\documentclass[a4paper,12pt]{article}
\pdfoutput=1
\usepackage{jheppub}

\usepackage{graphics}      
\usepackage{graphicx}
\usepackage{longtable}     
\usepackage{caption}
\usepackage{setspace}
\usepackage{amsmath}
\usepackage{bm}            
\usepackage{verbatim}
\usepackage[colorlinks=true]{hyperref}  

\newcommand\descitem[1]{\item{\bfseries #1}\\}

\title{Probing Non-Standard Neutrino Interactions with Supernova Neutrinos at Hyper-K}
\author{Minjie Lei,}
\author{Noah Steinberg,}
\author{James D. Wells}

\affiliation{Leinweber Center for Theoretical Physics, Department of Physics, University of Michigan
Ann Arbor, MI 48109 USA}

\abstract{Non-standard neutrino self interactions (NSSI) could be stronger than Fermi interactions. We investigate the ability to constrain these four-neutrino interactions by their effect on the flux of neutrinos originating from a galactic supernova. In the dense medium of a core collapse supernova, these new self interactions can have a significant impact on neutrino oscillations, leading to changes at the flavor evolution and spectra level. We use simulations of the neutrino flux from a 13 solar mass, core collapse supernova at 10 kpc away, and numerically propagate these neutrinos through the stellar medium taking into account vacuum/MSW oscillations, SM $\nu-\nu$ scattering as well as $\nu-\nu$ interactions that arise from NSSI. We pass the resulting neutrino flux to a simulation of the future Hyper-Kamiokande detector to see what constraints on NSSI parameters are possible when the next galactic supernova becomes visible. We find that these constraints depend strongly on the neutrino mass hierarchy and if the NSSI is flavor-violating or preserving. Sensitivity to NSSI in the normal hierarchy (NH) at Hyper-K is limited by the experiment's ability to efficiently detect $\nu_{e}$, but deviations from no NSSI could be seen if the NSSI is particularly strong. In the inverted hierarchy (IH) scenario, Hyper-K can significantly improve constraints on flavor-violating NSSI down to $\mathcal{O}(10^{-1})G_{F}$.}

\notoc 
\begin{document}
\maketitle
\flushbottom

\section{Introduction}
The standard picture of three neutrino flavor oscillations has become well established in the last 20 years, with many details, however, yet to be determined. In this picture, neutrinos behave according to the SM supplemented by some yet to be determined mass generation mechanism. Though this picture is rather complete, there is room for new physics to appear, among other ways, in the form of non-standard neutrino interactions (NSI), i.e. higher dimensional operators involving neutrinos and matter. NSI of the form
\begin{equation} \label{eq:1}
 \sqrt{2}G_{F}\varepsilon_{\alpha\beta}(\bar{\nu_{\alpha}}\gamma^{\rho}\nu_{\beta})(\bar{f}\gamma_{\rho}f),
 \end{equation}
 where $f = e,u,d$ and $\varepsilon_{\alpha\beta}$ is a $3\times3$ matrix of couplings in flavor space, have recently been the focus of intense scrutiny \cite{Rius:2018cjf,Experiments,Altmannshofer:2018xyo,Status}. In long baseline oscillation experiments (T2K, IceCube, DUNE, etc.)\ these non-standard interactions modify the matter potential through which neutrinos traverse and can greatly obscure the extraction of neutrino oscillation parameters ($\sin^2(2\theta_{ij})$, $\delta_{\text{CP}}$).

In contrast, somewhat less attention has been paid to non-standard $\textit{self}$ interactions (NSSI). In the Standard Model, neutrino self interactions occur through $Z$ boson exchange with an effective Hamiltonian given by:
\begin{equation} \label{eq:2}
\mathcal{H}_{SM}^{\nu-\nu} = \frac{G_{F}}{\sqrt{2}}\sum_{l,l' = e,\mu,\tau} (\bar{\nu}_{lL}\gamma_{\alpha}\nu_{lL})(\bar{\nu}_{l'L}\gamma^{\alpha}\nu_{l'L}),
\end{equation}
where $G_{F}$ is the Fermi constant. Because of the difficulty of directly measuring neutrino-neutrino scattering, NSSI of the schematic form $F(\bar{\nu}\gamma_{\alpha}\nu)(\bar{\nu}\gamma^{\alpha}\nu)$ where $F \gg \frac{G_{F}}{\sqrt{2}}$ is possible. Indirect bounds on $F$ were obtained from the decays $\pi^{+}\rightarrow e^{+}\nu_{e}\bar{\nu}\nu$ and $K^{+}\rightarrow l^{+}\nu_{l}\bar{\nu}\nu(l = e/\mu)$ \cite{Manohar:1987ec,Bilenky:1980ym} as well as from comparing the expected diffusion time of neutrinos exiting SN1987A with the time period over which the neutrinos were detected \cite{BialynickaBirula:1964zz}. These lead to bounds of roughly $F< (10^{3}-10^{7}) G_{F}$ where the lower bounds come from SN1987A and the higher bounds from meson decay. The tightest bounds on $F$ come from the 1-loop contributions of the above NSSI operator to the invisible width of the $Z$ boson \cite{Bilenky:1994ma}, which is known to $\sim 0.3\%$ \cite{Tanabashi:2018oca}. Here, the new 1-loop amplitude interferes with the SM amplitude with the same final state, and the authors obtain roughly $F<\mathcal{O}(10)G_{F}$. Clearly NSSI with an interaction strength greater or equal in magnitude to the Fermi interaction is still allowed from these results.
\vskip 0.15in
Recently, it has been understood that the extremely high densities of core collapse supernova, in conjunction with the relatively large neutrino flux, could be an ideal environment to look for NSSI \cite{Raffelt:2013rqa,Blennow:2008er,Nakazato:2012qf,Dighe:2017sur,Das:2017iuj,Duan:2006an,Duan:2008eb}. With the construction of the next generation of large scale neutrino oscillation experiments (DUNE, Hyper-K), the next occurence of a galactic supernova could give access to NSSI parameters $\leq\mathcal{O}(10^{-1})G_{F}$ if it is less than $\sim10$ kpc away. Analytic as well as large scale computational techniques have been developed both to study the effects of standard neutrino oscillations \cite{Duan:2006an,Duan:2008eb,Fogli:2004ff,Dighe:1999bi,Doring:2019axc,Shalgar:2019kzy}, and to look for features of NSSI at the flavor evolution and spectra level \cite{Dighe:2017sur,Kharlanov:2019cpk}. However, most of the previous works use schematic neutrino flux data that have relatively large spectra difference between flavors, while realistic supernova neutrino data during the relevant emission periods are often far more degenerate, suppressing oscillation features. Moreover, although there has been significant progress in next-generation neutrino detectors since the first detection of SN1987A, the actual number of neutrino events that can be detected from a future SN is still limited due to the extremely weak interactions of neutrinos. Thus, some features of NSSI at the supernova emission level might not be preserved at the detector data level, a concern we wish to explore.

	In this paper we build on previous work by applying computational techniques to neutrino flux from a realistic numerical supernova simulation \cite{Nakazato:2012qf}, and look for NSSI signatures by simulating detection data at the Hyper-K detector \cite{Abe:2018uyc}. We develop simple physical observables to examine in the event of future galactic supernova and estimate the sensitivity of Hyper-K to NSSI after applying these more realistic simulations of neutrino flux and detector response.

\section{Neutrino Production in Core Collapse Supernova}
\begin{figure}[t]
\begin{center}
\includegraphics[width=16cm]{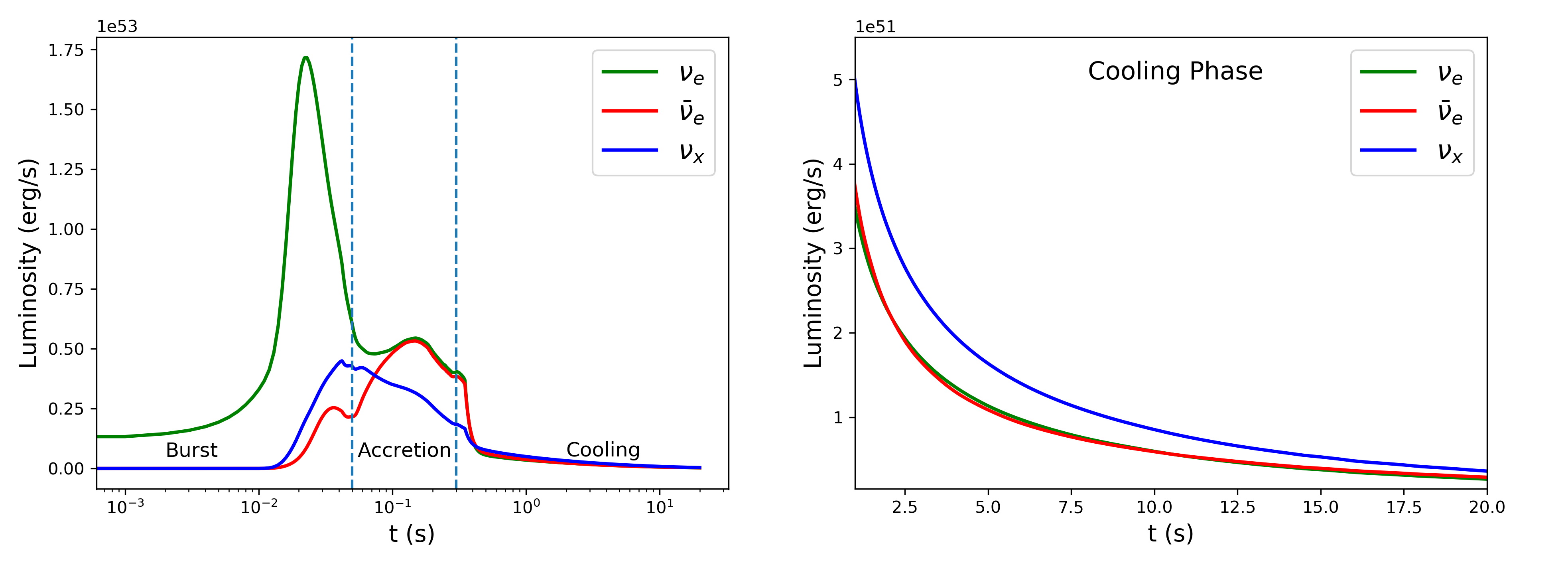}
\caption{Plots of luminosity over time from reproducing results from Nakazato et al.\ \cite{Nakazato:2012qf} for a model with initial stellar mass $13m_{\odot}$, metallicity 0.02, and shock revival time 300 ms. The left panel corresponds to the whole time profile from infall to 20 seconds after core bounce. The right panel gives a zoomed in look at the luminosity evolution during the later cooling stage. Since the luminosity of $\nu_{\mu}, \bar{\nu}_{\mu}, \nu_{\tau}$, and $\bar{\nu}_{\tau}$ are degenerate, they are denoted collectively as $\nu_{x}$.}
\label{fig:flux_sim}
\end{center}
\end{figure}

We begin with a brief description of the current picture of neutrino production in a core collapse supernova. When a sufficiently massive star ($M > 8M_{\odot}$) can no longer sustain itself through nuclear fusion, its core collapses to a proto-neutron star, ejecting its stellar envelope and setting off the supernova explosion. Neutrinos are produced in three distinct stages:

\begin{enumerate}
\descitem {Initial collapse and neutronization burst of neutrinos (0 $< t <$ 100 ms)} As the star runs out of nuclear fuel, its core is supported mainly by degenerate electrons. As the core contracts, these electrons are captured by nuclei and free protons, producing a large flux of $\nu_{e}$. The reduction in degeneracy pressure further collapses the core, leading to a runaway production of $\nu_{e}$ which stream out of the star. When the density of the inner core approaches $3\times10^{11}\ \text{g}/\text{cm}^{3}$ the neutrinos become trapped in the dense matter \cite{Nbook}. The core stops collapsing and rebounds, sending a shockwave outwards, dissipating energy by photodissociation of nuclei leaving behind a plethora of free nucleons. These nucleons capture electrons, producing a large number of $\nu_{e}$ which pile up behind the shock wave, until they reach a zone of low enough density and are released in a few milliseconds. This is known as the neutronization burst.
\descitem{Shockwave stall and thermal neutrino flux (100 ms $< t \lesssim$ 300 ms)} In the next phase of neutrino production, the shockwave stalls and matter accretes on the proto-neutron star and heats up, leading to the thermal production of neutrinos and anti-neutrinos of all flavors. The low matter density outside the core means this thermal flux of neutrinos can freely stream out of the star, corresponding to a period of neutrino emission dubbed the accretion phase.
\descitem{Helmholtz cooling phase and degenerate neutrino flux (300 ms $< t$)} After the shockwave is revived the explosion sets in, with accretion by the proto-neutron star stopping shortly after. The star enters a Kelvin-Helmholtz cooling phase, during which it emits neutrinos and antineutrinos of all flavors on a time scale of $\sim 10$ seconds.
\end{enumerate}
The above serves as a qualitative picture of neutrino production and emission which depends on the supernova model used. For a more detailed overview of neutrino production in supernova, see Mirizzi et al.\ \cite{Mirizzi:2015eza}. In Fig.~\ref{fig:flux_sim} we show the luminosity of supernova neutrinos vs.\ time from a 13 solar mass star using simulations by Nakazato et al.\ \cite{Nakazato:2012qf}. These simulations neither include NSSI nor take into account subsequent neutrino oscillations (implementation of which we make in this paper), but instead give the approximate time evolution of the neutrino flux as the core collapse proceeds. The left plot of Fig.~\ref{fig:flux_sim} covers the whole time profile from infall to 20 seconds where the initial spike of $\nu_{e}$ from the neutronization burst is prominent, followed by an increase in flux of $\nu_{x}$ ($x$ is any other (anti)lepton flavor) over the accretion phase. The right plot shows the cooling phase from $\sim 1$ second on, showing the near degeneracy of the flux between all flavors.
\section{NSSI and Neutrino Transport in Supernova Simulations}
The dynamics of neutrino transport in supernova is a non-linear problem involving three distinct oscillation contributions:
\begin{enumerate}
\item Vacuum oscillations, which depend on the splittings between mass eigenstates and their relative ordering,
\item The MSW effect, which can cause fast flavor conversion in the presence of extremely large matter densities,
\item Forward elastic scattering between neutrinos. This leads to non-linear `collective oscillations' that couple neutrinos of different energies, and produce bipolar oscillations that split and/or swap the energy spectrum of neutrinos of different flavors.
\end{enumerate}
The evolution of the neutrino flavor ensemble in the presence of self interactions was first described by Sigl and Raffelt \cite{Sigl:1992fn}. The equation of motion for each neutrino and antineutrino momentum mode is given by
\begin{align}
i\dot{\rho}_{p} &= [(\Omega^{0}_{p} + V + \Omega^{S}_{p}), \rho_{p}], & -i\dot{\bar{\rho}}_{p} &= [(\Omega^{0}_{p} - V - \Omega^{S}_{p}), \bar{\rho}_{p}].
\end{align}
Here $\rho_{p}(\bar{\rho}_{p})$ is the density matrix for neutrinos(anti-neutrinos) of momentum $p$ in the flavor basis. The diagonal elements give the occupation numbers of each neutrino flavor and the off-diagonal elements contain correlations between the mixing flavors \cite{Sigl:1992fn}. For ultra-relativistic neutrinos, $\Omega^{0}_{p} = U\text{diag}(m^{2}_{1}, m^{2}_{2}, m^{2}_{3})U^{\dagger}/2p$ is the free Hamiltonian ($U$ is the PMNS matrix),  $V = \sqrt{2}G_{F}n_{e}\text{diag}(1,0,0)$ in the weak interaction basis gives the contribution to forward scattering off of electrons, and the last term, $\Omega^{S}_{p}$, is the additional contribution to scattering from self interactions:
\begin{equation}
\Omega^{S}_{p} = \sqrt{2}G_{F}\int d^{3}q (1 - \vec{v}_{q} \cdot \vec{v}_{p})\{G(\rho_{q} - \bar{\rho}_{q})G + GTr[(\rho_{q} - \bar{\rho}_{q})G]\}.
\end{equation}
From here on we work in the 2 x 2 flavor space of $\nu_{e}$ and $\nu_{x}$, where $x = \mu, \tau$. In the SM, the $G$ matrix is the identity. In the presence of NSSI, after parameter redefinitions discussed in \cite{Dighe:2017sur}, $G$ takes the form
\begin{equation}
    G = \begin{bmatrix}
        1 + g_{3} & g_{1} \\
        g_{1} & 1 - g_{3}
     \end{bmatrix},
\end{equation}
where $g_{1}$ contibutes to flavor-violating NSSI (FV-NSSI), and $g_{3}$ contributes to flavor-preserving NSSI (FP-NSSI). Evaluating what possible constraints can be placed on these two couplings with Hyper-K is a central goal of this paper.
\vskip 0.15in
In the supernova environment, there are two broad regimes of neutrino transport and oscillation: the early high matter density regime where the neutrino mean free paths are less than the neutrino oscillation lengths, and the late coherent regime where neutrinos stream out freely from the proto-neutron star. In the first regime, the high matter density suppresses collective neutrino oscillations \cite{Chakraborty:2011gd}, which only become relevant when the neutrino number density is comparable to or greater than the nucleon number density. Since NSSI modify collective neutrino interactions, this first regime is less discerning of NSSI, and so we turn to the second regime. For the numerical supernova simulations used in this paper \cite{Nakazato:2012qf}, we identify the second regime as roughly 1 second after core bounce, when the inner core of the progenitor star has settled into a proto-neutron star of radius $\sim$ 10km, from which neutrinos stream out freely, until 20 seconds, when the flux has significantly dropped off. This period, from $1$ to $20$ seconds, is the most sensitive to NSSI, so it will be the focus of this paper.

To simulate neutrino flavor evolution in the coherent regime ($t > 1\text{ s}$), we follow the computational approach developed by Duan et al.\ \cite{Duan:2008eb}. Several approximations are made in this approach. First, neutrino emission during this regime follows the ``neutrino bulb model,'' in which neutrinos are emitted isotropically outwards from the surface of the proto-neutron star. A simplified supernova matter profile for the late time is assumed, which only depends on the distance from the center of the neutron star. Finally, in this paper, to allow the calculation of neutrino emission and evolution in the entire late time period from 1 to 20 seconds after core bounce, we use the single angle approximation instead of the multi-angle approximation used in \cite{Duan:2008eb}. In this approximation, all neutrinos evolve in the same way as a radially-propagating neutrino. In the real supernova environment, there are a variety of physical processes that can affect neutrino transport and evolution, complicating the simplified procedure adopted here. However, we justify our approximations by reasoning that the effect of coupled neutrino oscillations and their modification by NSSI will always be present in any supernova environment, and the results from Duan et al.\ \cite{Duan:2006an,Duan:2008eb} as well as our own calculations have given us confidence that most of the qualitative and even the quantitative results are likely to survive in a more advanced supernova simulation. If more sophisticated supernova models become available, such as the complete evolution of the matter density profile, they can always be incorporated into the same approach to modify our results, with qualitative features and approximate sensitivity estimates expected to be only lightly touched.

\section{Computational Approach and Simulation Results}
Our computational approach is organized as follows. First, we apply the neutrino flavor transformation calculation procedure by Duan et al.\ \cite{Duan:2008eb} to neutrino emission at a particular point during the late time regime of the numerical supernova simulation, chosen to be 5 seconds after core bounce, and compare the standard oscillation results with those obtained by Duan et al.\ using schematic supernova neutrino flux. The supernova model used has initial mass $M_{\text{init}} = 13 M_{\odot}$ and metallicity $Z = 0.02$, and we use the same mixing angles and matter density profile (appropriately scaled to a star with $M_{\text{init}} = 13 M_{\odot}$), as those found in \cite{Duan:2008eb}. The choice of 5 seconds in our initial calculation is simply to confirm our simulation matches those found in the literature (it does) before extending it to the whole late time coherent regime. Then we incorporate the NSSI modification to collective oscillation and observe how different values of flavor-preserving ($g_{3}$) and flavor-violating ($g_{1}$) parameters alter the flavor evolution and neutrino spectra. Finally, we extend these calculations to the entire late time coherent regime from 1 to 20 seconds after core bounce, and feed the resulting post-oscillation neutrino flux to a simulation of the Hyper-K detector \cite{sntools} to look for signatures of NSSI in the detector data. The results of these calculations are presented in the following sections.

\subsection{Standard Oscillation - No NSSI}
\begin{figure}[t]
\begin{center}
\includegraphics[width=14cm]{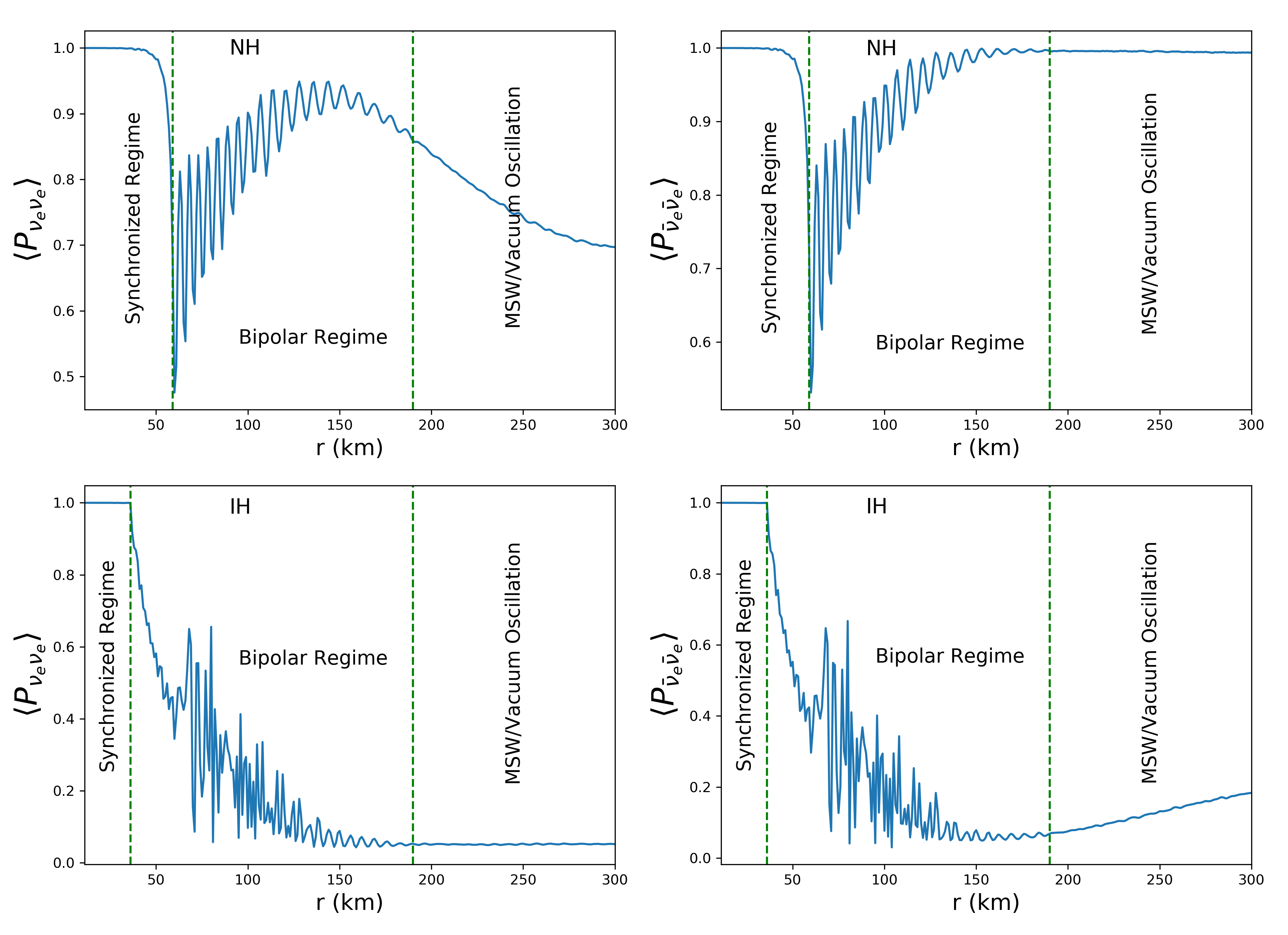}
\caption{Plots of energy-averaged neutrino survival probability evolution in the core-collapsed supernova environment. The upper panels correspond to the NH, lower panels to the IH, left panel $\nu_{e}$, and right panels $\bar{\nu}_{e}$. Three oscillation regimes can be identified from the plots. Near the proto-neutron star is the synchronized regime, where collective oscillations are suppressed by the large matter density, and neutrinos experience $\nu$-enhanced MSW flavor transformations. Far away from the star the neutrino flux is negligible and neutrinos undergo conventional vacuum/MSW oscillations. In between these two regions is the bipolar regime where neutrinos experience collective oscillations and spectral swaps/splits develop.}
\label{fig:fe_std}
\end{center}
\end{figure}

\begin{figure}[h]
\begin{center}
\includegraphics[width=16cm]{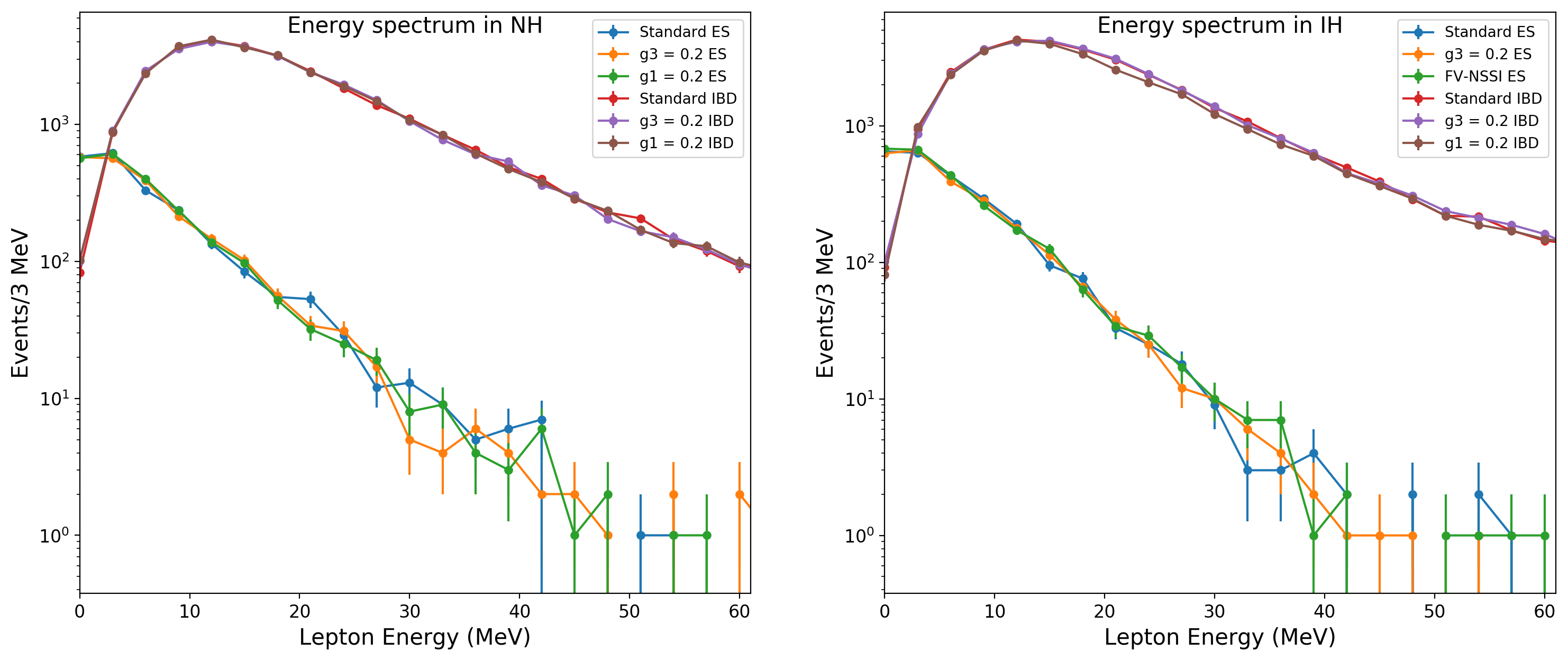}
\caption{Plots of visible lepton energy in Hyper-K from ES and IBD interactions. On the left is the NH and on the right is the IH. Figures include standard neutrino oscillations as well as contributions from FP-NSSI ($g_{1} \neq 0$) and FV-NSSI ($g_{3} \neq 0$). As can be seen, no splitting/swapping in the energy spectrum is visible.}
\label{fig:lepton_energy}
\end{center}
\end{figure}

Fig.~\ref{fig:fe_std} shows energy-averaged survival probability plots for the normal (NH) and inverted hierarchy (IH) for $\nu_{e}$ and $\bar{\nu_{e}}$. Three distinct oscillation regimes can be identified in the plots as a function of the radial distance from the neutrinosphere.  Our results agree with \cite{Duan:2008eb}, and show that collective oscillation phenomena only become important when the neutrino density becomes equal to or greater than the matter density ($n_{\nu} \geq n_{e}$), corresponding to the intermediate region of $50\text{ km} < r < 200\text{ km}$. At the energy spectra level, a spectral swap/split agreeing with \cite{Duan:2008eb} is observed. This swap/split is on the order of a couple of MeV. As can be seen in Fig. ~\ref{fig:lepton_energy}, the energy reconstruction limitations of Hyper-K \cite{Abe:2018uyc} make this behavior, which is only slightly modified by the addition of NSSI, largely unobservable. 

\subsection{NSSI Added}
\begin{figure}[t]
\begin{center}
\includegraphics[width=14cm]{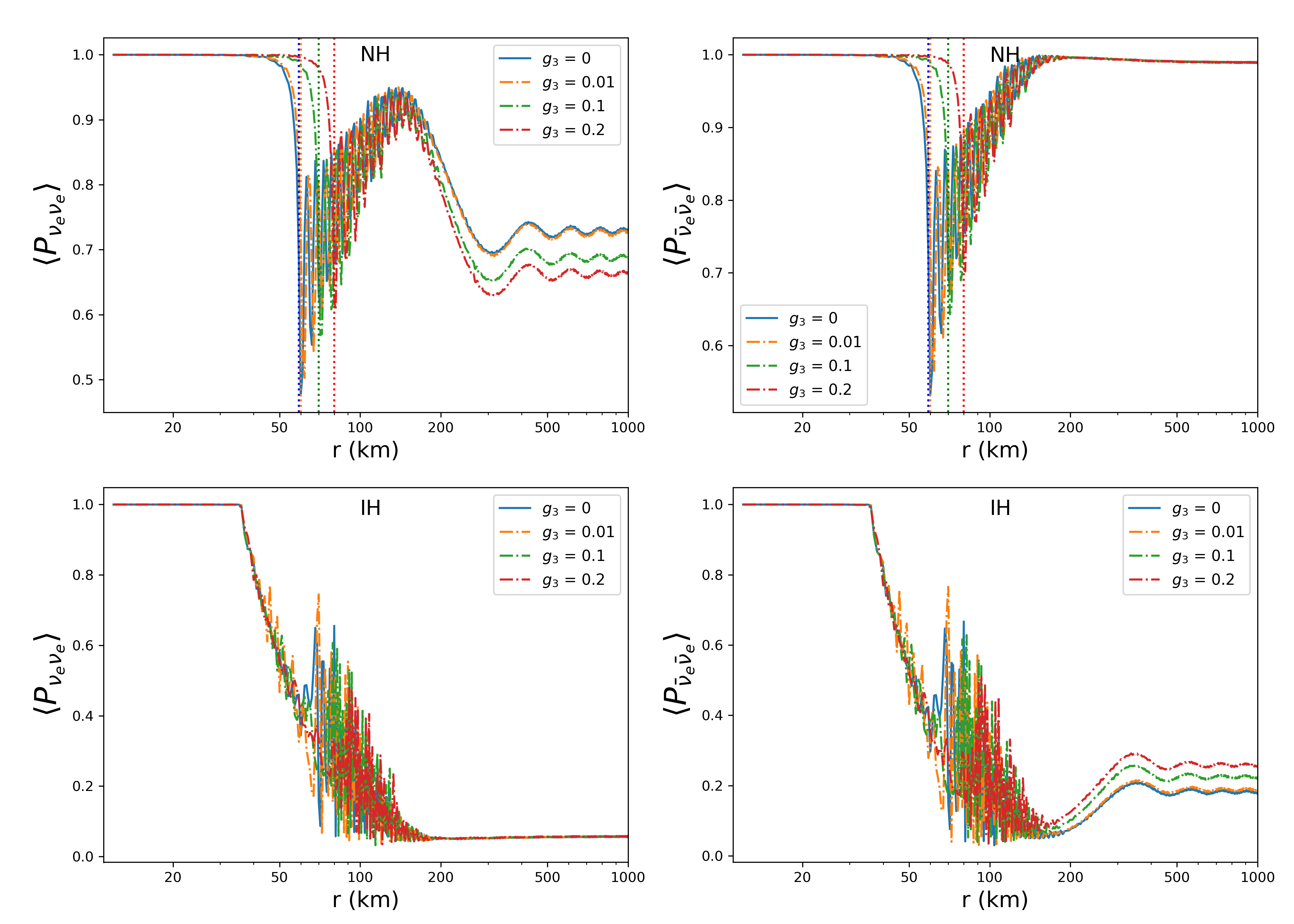}
\caption{Plots of energy-averaged survival probabilities with different values of FP-NSSI parameter $g_3$, the FV-NSSI parameter $g_1$ is set to zero for all cases. The panels are ordered as in Fig.~\ref{fig:fe_std} but with the x axis extended to 1000km, and put in log scale to focus in on the region where NSSI is most important. The x axis has also bFrom the plots we can see that larger values of $g_3$ correspond to a delay of the onset time of the bipolar regime in the NH case, as indicated by the vertical lines. The effect is less apparent in the IH case. Small differences in the final survival probability develop between different values of $g_3$ for $\nu_e$ in the NH case, and for $\bar \nu_e$ in the IH case.}
\label{fig:fe_fp}
\end{center}
\end{figure}

\begin{figure}[t]
\begin{center}
\includegraphics[width=14cm]{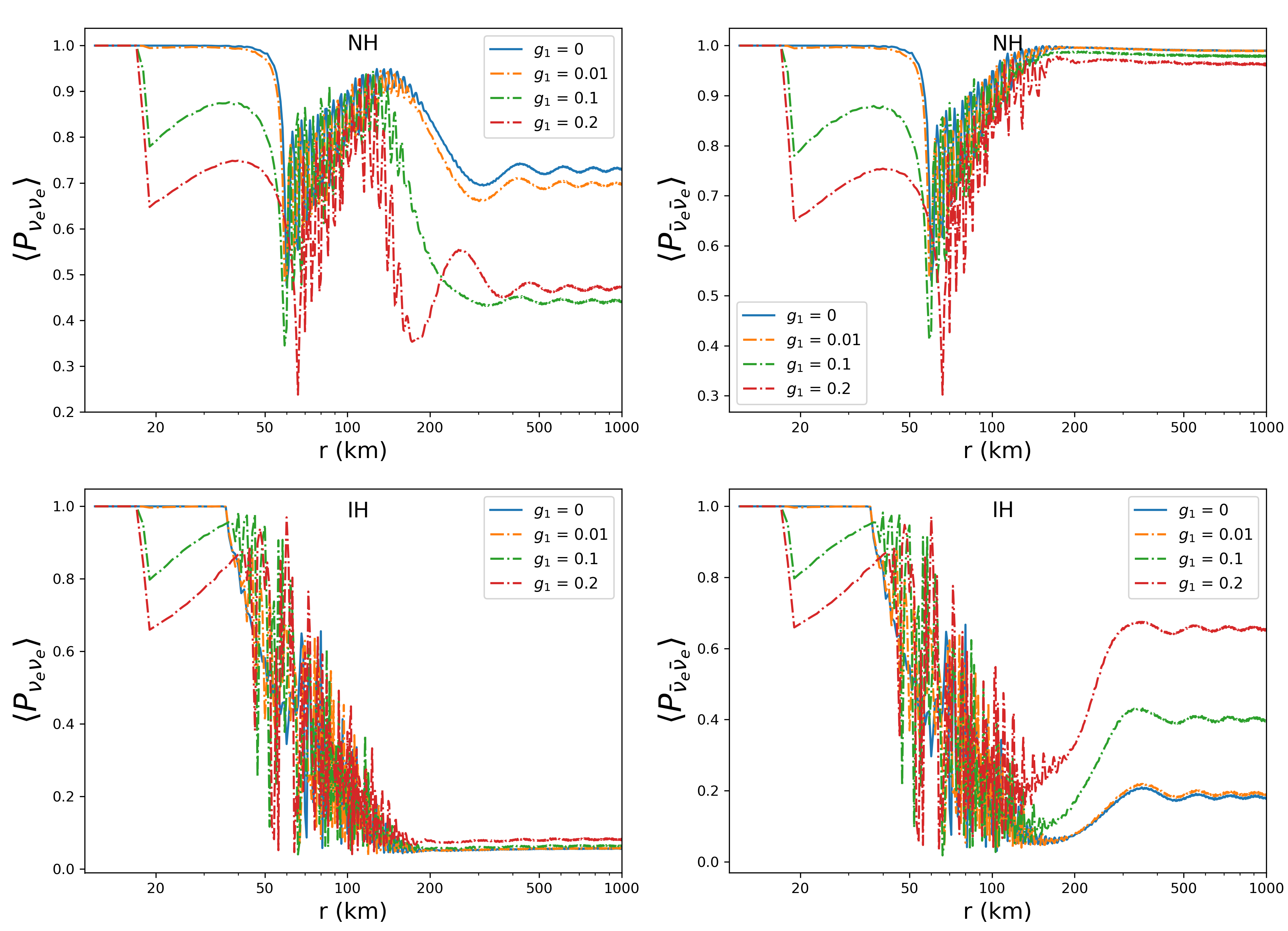}
\caption{Plots of energy-averaged survival probabilities with different values of flavor-violating NSSI parameter $g_1$, the flavor-preserving NSSI parameter $g_3$ is set to zero for all cases. The panels are ordered as in Fig.~\ref{fig:fe_std} but with the x axis extended to 1000km, and put in log scale to focus in on the region where NSSI is most important.Here, larger values of $g_1$ correspond to an advance of the onset time of the bipolar regime in the NH case. Large differences can be observed in the final survival probability between different values of $g_3$ for $\nu_e$ in the NH case, and for $\bar \nu_e$ in the IH case.}
\label{fig:fe_fv}
\end{center}
\end{figure}

Now we study the effects of NSSI by scanning $g_1$ and $g_3$ over the values \{0.01, 0.1, 0.2\}. We turn on only one NSSI parameter at a time to more accurately understand their individual effects on the neutrino spectra. Figs.~\ref{fig:fe_fp} ($g_{3}$ turned on) and~\ref{fig:fe_fv} ($g_{1}$ turned on) show the energy-averaged survival probability as a function of distance from the neutrinosphere for the NH and IH and for $\nu_{e}$ and $\bar{\nu}_{e}$.

In Fig.~\ref{fig:fe_fp} we see that in the NH (top), larger values of $g_{3}$ tend to delay the onset of collective oscillation effects, while for the IH (bottom) the delay disappears. The final survival probability is slightly modified for $\nu_{e}$ in the NH (top left) and $\bar{\nu}_{e}$ in the IH (bottom right), with no change in final survival probability for the opposite cases.

In Fig.~\ref{fig:fe_fv} we see the flavor-violating coupling, $g_{1}$, has a much more dramatic impact on the survival probabilities. Larger values of $g_{1}$ tend to advance the onset of collective oscillations, as can be seen in all four plots, as well as substantially modify the final survival probability for $\nu_{e}$ in the NH (top left) and $\bar{\nu}_{e}$ in the IH (bottom right), with little to no change in final survival probability for the opposite cases.

An analytical and qualitative explanation for this behavior can be found in \cite{Dighe:2017sur}, where the authors obtain a linearized, analytic formula for the rate of flavor conversions, represented by $S_{\omega,u}$, where $\omega$ is the neutrino energy and $u = sin^{2}(\theta_{R})$ represents the emission angle. For only one non-zero NSSI parameter these equations of motion are eigenvalue equations with solutions that grow like $e^{\kappa t}$ for $\kappa$ real. For FP-NSSI (FV-NSSI) one finds that $\kappa$ is reduced (enhanced) by a factor of $1-g_{3}^{2}$ ($1+g_{1}^{2}$) relative to the no-NSSI case, which leads to the delay (advancement) of collective oscillation effects. Thus, our results agree with \cite{Dighe:2017sur}.

As our goal is to understand the effects of NSSI on the observed spectrum at Hyper-K of a future galactic supernova, we are most interested in the differences between the energy-averaged final survival probabilities for different NSSI parameter values. The final flux of electron neutrinos ($\nu's$ per $4\pi$ steradian per second), $F^{f}_{e}$, is given in terms of the initial fluxes $F_{e}, F_{x}$ and the survival probability $P_{\nu_{e}\nu_{e}}$ (the survival probability at the surface of the star)  by

\begin{equation}
F^{f}_{e} = P_{\nu_{e}\nu_{e}}F_{e} + (1-P_{\nu_{e}\nu_{e}})F_{x} = P_{\nu_{e}\nu_{e}}(F_{e} - F_{x}) + F_{x},
\end{equation}
with an analogous formula for anti-electron neutrinos. During the late-time cooling phase, where NSSI have the largest impact, the flux hierarchy is $F_{x} > F_{e},\bar{F}_{e}$ \cite{Nakazato:2012qf}. Therefore, larger values of $P_{\nu_{e}\nu_{e}}$ correspond to a smaller final flux. Examining Figs.~\ref{fig:fe_fp} and~\ref{fig:fe_fv}, we can predict that for $\nu_{e}(\bar{\nu}_{e})$ in the NH(IH), the final flux in the flavor-violating NSSI scenario will be larger(smaller) than the final flux in both the flavor-preserving NSSI and standard scenarios. As the cooling phase continues, the fluxes of different flavors become more and more degenerate and this difference approaches zero.

\subsection{Hyper-K Sensitivity to NSSI}
In this section we discuss the detection capabilities of Hyper-K. The peak of the supernova neutrino energy spectrum is around  $10-15$ MeV, with an exponentially falling tail. Only three interaction channels have low enough thresholds at these energies: inverse beta decay (IBD) $\bar{\nu}_{e} +\,p \rightarrow e^{+} +\,n$, neutrino-electron elastic scattering (ES) $\nu + e^{-} \rightarrow \nu + e^{-}$, and charged current scattering off of oxygen-16 nuclei (OCC) $\nu_{e}(\bar{\nu}_{e}) + ^{16}O \rightarrow e^{-}(e^{+}) + X$. The initial burst of neutrinos from the neutronization phase of the supernova will mostly be $\nu_{e}$, which will allow for an unambiguous identification of the supernova onset time to within 1 ms \cite{Abe:2018uyc}. At Hyper-K, we expect the event distribution to be dominated by IBD, resulting in 50,000 to 75,000 events, along with 3,000 to 4,000 ES and 1,000 to 14,000 OCC events detected over a 20 second period for a core collapse supernova at 10 kpc. Roughly 70 - 80\% of these events occur in the 1 to 20 second time period of interest, as can be gleaned from \cite{Abe:2018uyc}. Looking at Fig.~\ref{fig:fe_fv}, this is optimal for constraining flavor-violating NSSI in the IH, as the $\bar{\nu}_{e}$ final survival probability depends strongly on the value of $g_{1}$. In the NH, the final $\nu_{e}$ survival probability also depends strongly on $g_{1}$ but poor statistics in the ES channel worsen the ability of Hyper-K to distinguish between NSSI and no NSSI.

\begin{figure}[t]
\begin{center}
\includegraphics[width=14cm]{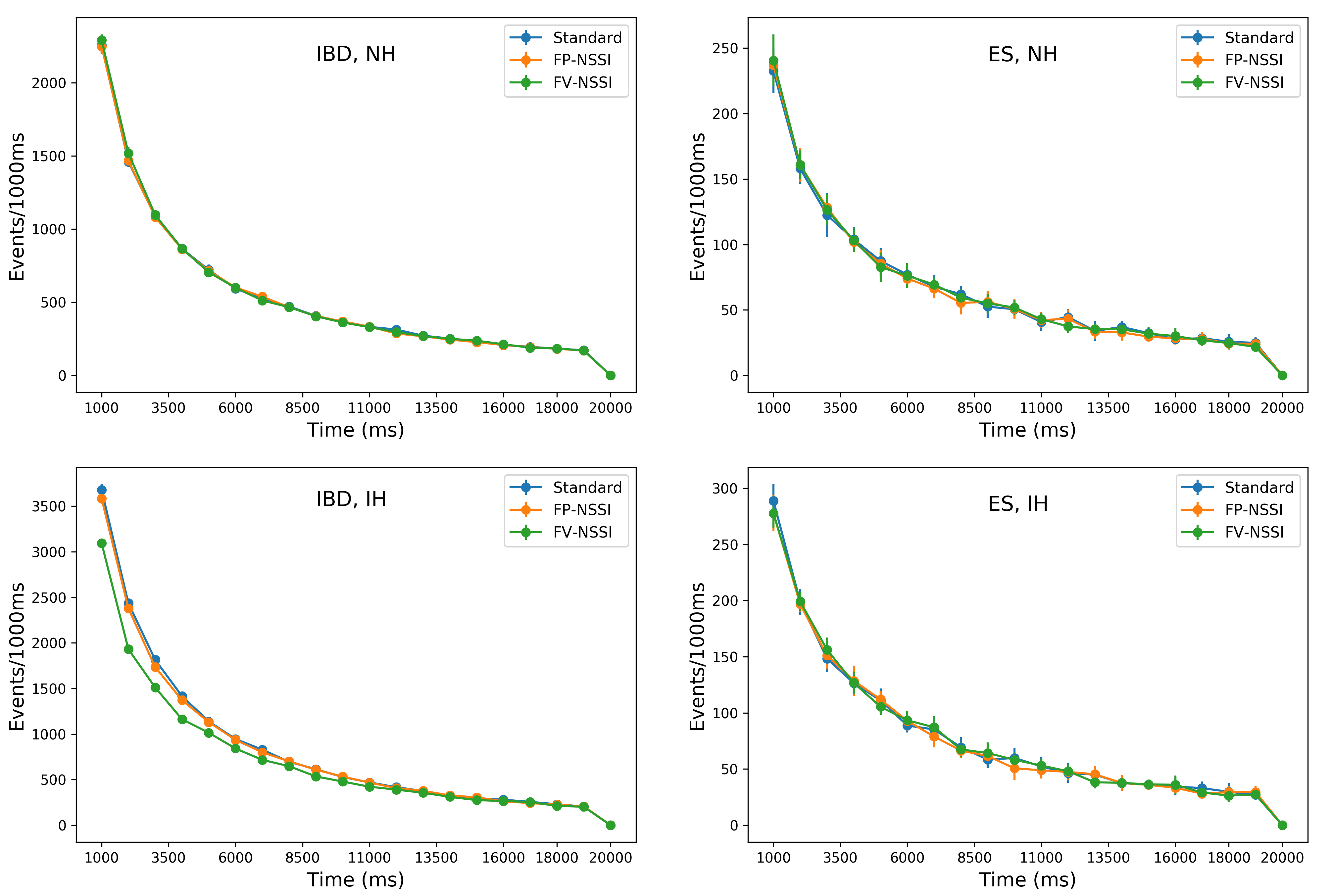}
\caption{Plot of number of detected events over time, starting at $t = 1$ second. $t = 0$ is set to the time of the observed onset of the neutrino flux. The upper panels correspond to NH, lower panels IH, left panels the inverse beta decay channel, and the right panels the elastic scattering channel. Error bars are statistical only. Since the IBD channel correspond to $\bar{\nu_{e}}$ detection and ES channel mainly correspond to $\nu_{e}$ detection, as predicted by the flavor evolution plots in Figs.~\ref{fig:fe_fp} and~\ref{fig:fe_fv}, we should expect to see significant differences between flavor-violating NSSI and flavor-preserving NSSI/standard for IBD in the IH case, and for ES in the NH case, and the difference should decrease as neutrino fluxes become more degenerate over time. This is clearly the case for the IBD channel. However, it is not observed in the ES channel, due to poor $\nu_{e}$ statistics from water cherenkov detectors such as Hyper-K.}
\label{fig:detector_data}
\end{center}
\end{figure}

In Fig.~\ref{fig:detector_data} we show the number of IBD (left) and ES (right) events vs.\ time for both the NH (top) and IH (bottom), for no NSSI, $\{g_{1}, g_{3}\} = \{0, 0.2\}$ (FP-NSSI), and $\{g_{1}, g_{3}\} = \{0.2, 0\}$ (FV-NSSI). No corresponding figure is shown for the OCC events as their contribution to the overall event rate in the 1 to 20 second time window is at least an order of magnitude smaller than the ES and IBD event rates. It is easy to see that the IBD channel cannot easily constrain NSSI in the NH because the final survival probability for $\bar{\nu}_{e}$ (Figs.~\ref{fig:fe_fp} and~\ref{fig:fe_fv}) does not depend strongly enough on $g_{1}$ or $g_{3}$. Similarly, ES at Hyper-K cannot well constrain NSSI in either hierarchy, but this is mainly due to poor statistics in this channel. The largest difference can be seen in the bottom left plot of Fig.~\ref{fig:detector_data} in the IH, IBD channel, where there is a significant reduction in the event rate for FV-NSSI interactions with strength $g_{1} = 0.2$.

Based on numerical supernova simulations \cite{Nakazato:2012qf}, neutrino flux and luminosity during the cooling phase are largely independent of supernova models, so we should restrict any observable to this phase of the supernova (1 s $<$ $t$ $<$ 20 s), which fortuitously is the most sensitive to NSSI. Additionally, to remove the dependence on the distance of the supernova from the detector (which affects the absolute neutrino flux), and minimize the effect of systematic errors, one can take ratios of events in different interaction channels or in different time bins. 

In the standard Hyper-K setup with only water, it is difficult to distinguish between IBD and ES interactions on an event by event basis. This is because electrons and positrons are indistinguishable in a water cherenkov detector, but two solutions are possible. ES and IBD events may be separated on a statistical basis by the angular distributions of the produced (anti-)lepton, as electrons from ES will be approximately forward going and positrons produced from IBD will be essentially isotropic. The second option is neutron tagging which the EGADS collaboration is currently investigating by doping water with Gadolinium \cite{Xu:2016cfv}. If Gadolinium (Gd) is added to the water in Hyper-K then IBD interactions are identifiable via the resultant neutron capture. Gadolinium has a substantial thermal neutron capture cross section of 49,000 barns and emits an 8 MeV gamma cascade which can be detected via its cherenkov light. At a concentration of 0.1-0.2\% the efficiency for neutron capture is roughly 90\%. 

A combination of statistical, cut based methods as well as the addition of gadolinium could allow for IBD identification with an efficiency between $90 - 100\%$. In this case we identify an observable, $R_{\text{IBD}}$\eqref{eq:ratio}, the ratio of total ES + OCC events to total IBD events from $1\text{ s} < t < 20\text{ s}$ after the onset of the neutrino flux, to be an indicator of NSSI. 

\begin{equation}
R_{\text{IBD}} = \frac{\text{ES and OCC events from 1 s $<$ $t$ $<$ 20 s }}{\text{IBD events from 1 s $<$ $t$ $<$ 20 s}},
\label{eq:ratio}
\end{equation}
where $t = 0$ is the standardly defined onset time of the neutrino flux.

Assuming $100\%$ IBD identification efficiency, we show results for $R_{\text{IBD}}$ in Table \ref{tab:evt_ratio} where the errors on each ratio are statistical. FV-NSSI in the IH with $g_{1} = 0.2$ or greater will lead to an increase in this ratio of more than $10\%$, which we estimate to be discernible from the no NSSI standard scenario. For IBD identification below $100\%$ efficiency, the discerning power of this observable rapidly deteriorates. This is because the number of IBD events is an order of magnitude larger than ES and OCC events, so any migration of IBD events due to misidentification will quickly change $R_{\text{IBD}}$. Below in Fig.~\ref{fig:ibd_eff} we show the sensitivity of $R_{\text{IBD}}$ to the IBD identification efficiency by plotting this observable for FV-NSSI in the IH as a function of the IBD identification efficiency. As the IBD identification efficiency drops below $99\%$ $R_{\text{IBD}}$ rapidly deviates from its nominal value at $100\%$ efficiency making NSSI vs no NSSI distinctions difficult. Thus, this observable becomes less reliable as the IBD identification efficiency deteriorates.

\begin{table}[t]
\caption*{$R_{\text{IBD}}\times 100\%$}
\centering
\scalebox{1}{
\begin{tabular}{|c|c|c|c|}
\cline{2-4}
\multicolumn{1}{l|}{} & Standard & FP-NSSI & FV-NSSI\\ \hline
NH & 12.95$\pm$0.37\% & 12.75$\pm$0.36\% &12.72$\pm$0.36\% \\
IH & 10.14$\pm$0.26\% & 10.21$\pm$0.26\% & 11.52$\pm$0.30\% \\
\hline
\end{tabular}}
\caption{$R_{\text{IBD}}$\eqref{eq:ratio}, the ratio of total events detected during the late time regime ($1\text{ s} < t < 20\text{ s}$) in the ES and OCC channels to those in the IBD channel for both NH and IH cases, assuming $100\%$ IBD identification efficiency. The ratio provides a largely model-independent signature for NSSI modifications. In this case, there is a significant difference between flavor-violating NSSI and flavor-preserving NSSI/standard in the IH case. With better ES channel statistics, such as from a liquid-Ar detector like DUNE, we expect to see much smaller statistical variation in the ratios, as well as a difference in the NH case, as predicted by the flavor history plots in Fig.~\ref{fig:fe_fp} \&~\ref{fig:fe_fv}.}
\label{tab:evt_ratio}
\end{table}

\begin{figure}[t]
\begin{center}
\includegraphics[width=11cm]{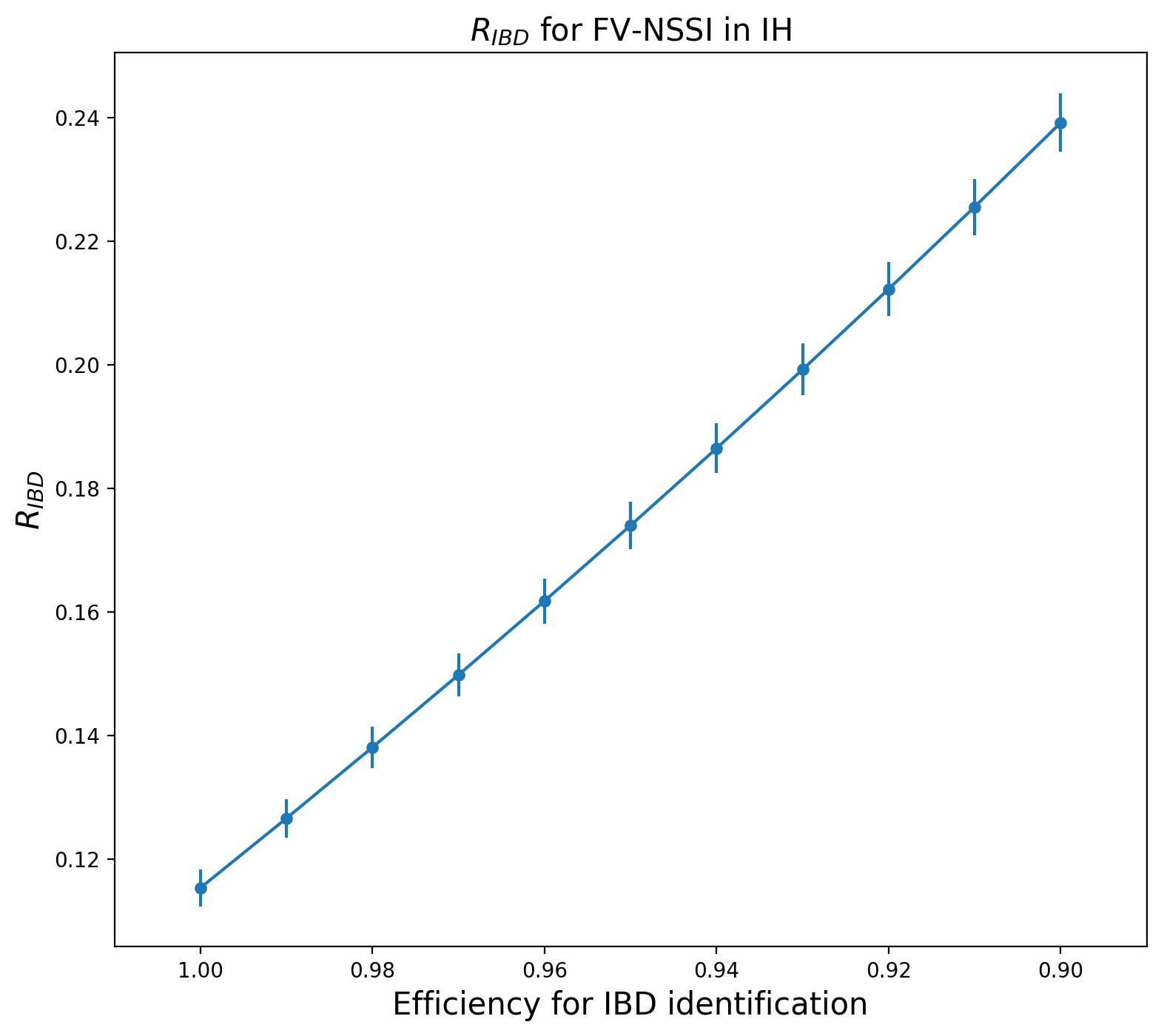}
\caption{Plot of $R_{\text{IBD}}$ for FV-NSSI in the IH vs IBD identification efficiency. As the IBD identification efficiency drops below $99\%$ $R_{\text{IBD}}$ rapidly deviates from its value at $100\%$ efficiency making NSSI vs no NSSI distinctions difficult. Thus, this observable becomes less reliable as the IBD identification efficiency deteriorates.}
\label{fig:ibd_eff}
\end{center}
\end{figure}

In the event that the IBD identification efficiency is not adequate for looking for the effects of NSSI, one should construct an observable that is independent of the ability of Hyper-K to identify specific neutrino interaction types. We take this observable, $R_{\text{10/20}}$, to be the ratio of the total number of events between 1 and 11 seconds to the total number of events between 11 and 20 seconds. 

\begin{equation}
R_{\text{10/20}} = \frac{\text{total events from 1 s $<$ $t$ $<$ 11 s }}{\text{total events from 11 s $<$ $t$ $<$ 20 s}},
\label{eq:time_ratio_eq}
\end{equation}
where $t = 0$ is the standardly defined onset time of the neutrino flux.

The behavior of this observable can be seen from Fig.~\ref{fig:detector_data}. For IBD in the inverted hierarchy, one can see that the Standard, FP-NSSI, and FV-NSSI event rates are nearly degenerate for $t > 11\text{ s}$, but the FV-NSSI event rate is suppressed in the region from $1\text{ s} < t < 11\text{ s}$. In the NH this suppression is not present because the final survival probability is only very weakly dependent on $g_{1}$ and $g_{3}$ as can be seen in Figs.~\ref{fig:fe_fp} and~\ref{fig:fe_fv}. We show results for $R_{\text{10/20}}$ in Table \ref{tab:time_ratio} where the errors on each ratio are statistical. FV-NSSI in the IH with $g_{1} = 0.2$ or greater will lead to an increase in this ratio of $10\%$, which again we estimate to be discernible from the no NSSI standard scenario. We note that this observable is not sensitive to the 20 second endpoint because statistics rapidly fall for $t > 20\text{ s}$.

\begin{table}[t]
\caption*{$R_{\text{10/20}}\times 100\%$}
\centering
\scalebox{1}{
\begin{tabular}{|c|c|c|c|}
\cline{2-4}
\multicolumn{1}{l|}{} & Standard & FP-NSSI & FV-NSSI\\ \hline
NH & 20.76$\pm$0.50\% &20.16$\pm$0.49\% &20.38$\pm$0.40\% \\
IH & 16.54$\pm$0.35\% & 17.14$\pm$0.36\% & 18.32$\pm$0.40\% \\
\hline
\end{tabular}}
\caption{$R_{\text{10/20}}$\eqref{eq:time_ratio_eq}, the ratio of total events detected during $1\text{ s} < t < 11\text{ s}$ in to those detected during  $11\text{ s} < t < 20\text{ s}$ for both NH and IH cases. This ratio provides a largely model-independent signature for NSSI modifications that is independent of Hyper-K's ability to identify the neutrino interaction type. In this case, there is a significant difference between flavor-violating NSSI and flavor-preserving NSSI/standard in the IH case.}
\label{tab:time_ratio}
\end{table}

\section{Conclusion}
Probing NSSI with supernova neutrinos presents a unique opportunity to search for new physics in the neutrino sector. By using a realistic supernova simulation under a suite of reasonable assumptions and combining it with advanced detector simulations, we have evaluated the ability of Hyper-K to constrain NSSI. We have shown that a single galactic supernova combined with Hyper-K's massive detection volume will allow one to constrain flavor-violating NSSI in the IH to two orders of magnitude smaller than current bounds ($g < 0.2$ vs.\ $g < 10$ previously) even if Hyper-K is unable to identify the neutrino interaction type. Additionally, a complementary experiment like DUNE, which uses liquid argon, will give a much better measurement of the $\nu_{e}$ flux from a supernova, allowing us to probe NSSI in the NH as well. Future work will focus on higher order improvements to the supernova simulation, such as implementing more realistic density profiles, full three flavor neutrino framework, and multi-angle instead of single angle approximation. We also wish to extend these results to DUNE after detector response characterization is more mature.

\acknowledgments

We thank Advanced Research Computing at the University of Michigan, Ann Arbor for their computational resources, as well as M. Buschmann, Z. Zhang, and K. Nakazato for helpful comments and discussion. The authors are supported by DoE grant de-sc0007859. The work of N. Steinberg was supported by a Leinweber Graduate Fellowship.

\bibliographystyle{utphys}
\bibliography{NSSI}

\end{document}